\documentclass[twocolumn]{aastex631}
\usepackage{amsmath}
\newcommand\p{\text{p}}
\newcommand{\rev}[1]{#1}
\newcommand{\revii}[1]{#1}

\begin{document}
\title{Measuring the Two-Dimensional Thermal Structures of Protoplanetary Disks}

\author[0000-0001-7155-3583]{Anna J. Fehr}
\affiliation{Center for Astrophysics \textbar~Harvard \& Smithsonian, 60 Garden St, Cambridge, MA 02138, USA}

\author[0000-0003-2253-2270]{Sean M. Andrews}
\affiliation{Center for Astrophysics \textbar~Harvard \& Smithsonian, 60 Garden St, Cambridge, MA 02138, USA}

\begin{abstract}
We present a flexible, annulus-by-annulus method to constrain the 2-D thermal structure of a protoplanetary disk from optically thick spectral line emission. Using synthetic disk models with a known temperature and density structure, we extracted the vertical emission surfaces and brightness temperatures in radial annuli for multiple CO isotopologue transitions and used them to infer the vertical temperature profiles. This approach reliably recovers the injected temperature structure despite noise and finite resolution. We demonstrated that even a modest set of emission lines can constrain the temperature across a wide range of radii and elevations. \rev{Nevertheless, biases in the extracted emission surfaces constitute a major source of systematic error.} Finally, we applied this method to archival ALMA observations of the HD 163296 disk, revealing that simple parametric radial temperature models may obscure the complexity of real disks and that additional observations are necessary to distinguish between different models of the vertical structure. This flexible framework can be readily applied to other systems, helping to characterize the thermal environments that shape planet formation.
\end{abstract}

\section{Introduction}
The thermal structures of protoplanetary disks are fundamental to both the physics and chemistry of disk evolution and planet formation. Physically, thermal structures set the local sound speeds and therefore the \rev{pressure scale height, a characteristic size scale in the disk that determines the vertical flaring} \citep{Kenyon87}. Moreover, they influence turbulent viscosities \citep{Shakura73} and the overall disk kinematics \rev{\citep[through the pressure gradient; e.g.,][]{Martire24, Longarini25}}. Chemically, these temperature distributions dictate the locations of sublimation fronts, or snow lines, where volatile species transition between gas and solid phases. These boundaries regulate the availability of icy materials, shaping the compositions of forming planets \citep{Oberg11, Oberg23}, and may serve as preferred sites for planet formation \citep{Stevenson88, Zhang15, Schoonenberg17}. 

Disk temperatures vary in the radial and vertical dimensions. The vertical gradient arises primarily from dust opacity, which regulates the transport of stellar irradiation through the disk \citep{Chiang97, DAlessio98}. The midplane remains relatively cool due to shielding by upper-layer dust, while the disk surface is exposed to direct stellar heating, resulting in a stratified thermal structure \citep{Calvet91}. However, the exact temperature distribution remains uncertain because dust opacity depends on grain size, composition, and number density -- parameters that are poorly constrained observationally and vary throughout the disk. Consequently, determining the true thermal structure requires additional measurements.

Efforts to determine the thermal structures of disks have generally followed two approaches. The first involves direct observational constraints on gas temperatures, using optically thick molecular line emission as a tracer \citep[e.g.,][]{Dartois03, Pinte18, Law21}. This method relies on identifying emission surfaces where brightness temperatures approximate the gas temperature. However, challenges arise from uncertainties in the elevations of emission surfaces, radiative transfer effects, and resolution limitations. The second approach employs thermochemical models to forward-model the observed line emission \citep[e.g.,][]{Woitke19, Calahan21}. While these models offer valuable insights, they depend on assumptions about dust opacity and dust and gas density distributions that may not fully capture real disk conditions.

In this work, we refine the former method by adopting a flexible approach in which radial annuli are treated independently. In each annulus, we measure the brightness temperatures and emitting layer heights for optically thick spectral lines and then infer the vertical thermal structure, $T(z)$. By testing our analysis on synthetic disk models, we evaluated the reliability of our technique and determined what additional observations could further improve temperature constraints. In Section \ref{sec:model}, we describe the physical conditions used in our model and the generation of synthetic image cubes. Section \ref{sec:methods} outlines how emission surfaces and radial temperature profiles were extracted and details the inference procedure. Section \ref{sec:results} explores systematic effects and limitations on precision. In Section \ref{sec:discussion}, we demonstrate our approach using archival ALMA data and discuss potential applications of our findings. Finally, Section \ref{sec:summary} summarizes the main conclusions.

\section{Disk Models and Synthetic Data}
\label{sec:model}
We generated synthetic disk observations to establish a controlled environment where the true thermal structure, $T(r, z)$, is known.  This enabled the systematic development and testing of a framework to extract temperature information from real ALMA data.  To achieve this, we constructed a disk model with physically motivated temperature and density distributions, then used radiative transfer calculations to produce synthetic image cubes. By comparing the recovered temperature structures to the known input model, we could assess the reliability and biases of the analysis techniques. The physical conditions of this model are summarized in Section \ref{sec:conditions}, and the process of translating the model into synthetic observations is detailed in Section \ref{sec:radtran}.

\subsection{Physical Conditions}
\label{sec:conditions}
The model treats the disk as an ideal gas in vertical hydrostatic equilibrium, where the density structure $\rho(r, z)$ balances gas pressure upwards and gravitational acceleration downwards \citep[e.g.,][]{Andrews24}.

\begin{figure}
    \centering
    \includegraphics{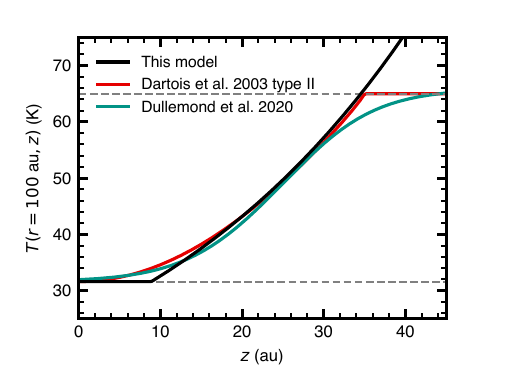}
    \caption{\rev{Example vertical temperature profiles following Equation \ref{eq:Tannulus} and the temperature profiles in \citet{Dartois03} and \citet{Dullemond20}. The dashed gray lines show $T_\text{mid}$ and $T_\text{atm}$ (used only in the \citealt{Dartois03} and \citealt{Dullemond20} profiles).}}
    \label{fig:tempprofile}
\end{figure}

We adopted a vertical temperature profile with an isothermal midplane layer that transitions to an exponential rise in the atmosphere,
\begin{equation}
\label{eq:Tannulus}
    T = \begin{cases}
        T_\text{mid} & \frac{z}{r} < a_\text{z} \\
        T_\text{mid} \exp[\gamma (\frac{z}{r}-a_z)] & \frac{z}{r} > a_z,
    \end{cases}
\end{equation}
\rev{where $T_\text{mid}$ is the midplane temperature, $a_z \, r$ is the height marking the transition from the isothermal midplane to the exponential rise, and $\gamma$ sets the slope of the rise. The temperature profile is} designed as a slight modification to the \citet{Dartois03} type I case that is more in line with detailed physical and thermochemical models \citep[e.g.,][]{Dalessio06,Qi19,Calahan21}. \rev{Unlike the \citet{Dartois03} type II and the \citet{Dullemond20} temperature prescriptions, this profile does not asymptotically flatten to an isothermal atmosphere temperature at some vertical layer. Consequently, an ``atmosphere temperature" is not well-defined (and thus we do not fit for such a parameter explicitly). The temperature profile used in this work is shown in Figure \ref{fig:tempprofile}, along with other temperature profiles commonly used in the literature.} 

Each of the vertical parameters varies with radius, such that
\begin{equation}
\label{eq:Tradial}
\begin{aligned}
    T_\text{mid} &= T_0\left(\frac{r}{r_0}\right)^{q_T} \\
    a_z &= a_0 \left(\frac{r}{r_0}\right)^{q_a} \\
    \gamma &= \gamma_0\left(\frac{r}{r_0}\right)^{q_\gamma}.
\end{aligned}
\end{equation}
We set $T_0 = 100$\,K, $q_T=-0.5$, $a_0=0.05$, $q_a=0.25$, $\gamma_0=5$, and $q_\gamma=-0.25$. \rev{These parameters were chosen to produce thermal structures comparable to the results from analogous modeling of real data \citep{Calahan21,Law21}.}
The surface densities followed the \citet{LyndenBell74} similarity solution,
\begin{equation}
    \Sigma = \Sigma_0\left(\frac{r}{r_0}\right)^{-p}\exp\left[-\left(\frac{r}{r_\text{t}}\right)^{2-p}\right],
\end{equation}
with $p=1$, $r_\text{t} = 160$\,au, and $\Sigma_0=100$\,g\,cm$^{-2}$.
All parametric normalizations used a reference radius of $r_0=10$\,au. 

\rev{The model neglects the effects of self‑gravity and radial pressure support when computing the gas velocity field. In other words, we have assumed that the disk kinematics are governed entirely by the gravitational potential of a 1 M$_\odot$ star. In reality, contributions from self-gravity and pressure support can alter that Keplerian rotation curve by modest amounts ($\sim$5\%; \citealt{Andrews24}). The details of the velocity structure enter into the measurements in Section \ref{sec:resolution}, and we quantify the effects of non-Keplerian contributions in Section \ref{sec:surfaceuncertainty}.}

We modeled the emission from rotational transitions of CO, $^{13}$CO, and C$^{18}$O, so we must also define the abundance distribution of CO and its isotopologues. We assumed a constant CO abundance relative to the total gas density, $X_\text{CO}=5\times10^{-5}$, within a vertical layer. \rev{This CO abundance is consistent with observational constraints \citep[e.g.,][]{Zhang21}, but likely represents a disk particularly abundant in CO.} The lower boundary of the layer was set by the CO freeze-out temperature, $T_\text{frz}=20$\,K \citep{vanZadelhoff01}, while the upper boundary, $z_\text{crit}$, was set by
\begin{equation}
    \sigma_\text{crit}=\int_\infty^{z_\text{crit}} \rho\,dz
\end{equation}
where the critical column density threshold, $\sigma_\text{crit} = 0.01$\,g\,cm$^{-2}$, approximates the photodissociation limit \citep{vanZadelhoff03, Aikawa06}.
Outside the layer, a depletion factor $D_\text{CO}=10^{-3}$ scales the CO abundance. \rev{This is consistent with estimates from thermochemical models \citep[e.g.,][]{Calahan21, PanequeCarreno25}, and the choice of depletion factor does not have a large effect on the relevant aspects of the emission morphology.}
We adopted fixed isotopic abundance ratios across the disk: CO$/^{13}\text{CO} =69$ and CO$/^{18}\text{CO}=557$ \citep{Wilson99}.

\subsection{Model Processing}
\label{sec:radtran}

Employing the prescription described above, we used the local thermodynamic equilibrium \rev{(LTE)} excitation and raytracing capabilities of {\tt RADMC-3D} \citep{Dullemond12} to generate synthetic spectral line cubes. The CO quantum properties were compiled from the LAMDA database \citep{Schoier05}, and line profiles were assumed to be Gaussian with widths set by thermal broadening. We simulated the $J$=1$-$0, 2$-$1, 3$-$2, 4$-$3, and 6$-$5 emission lines for CO, $^{13}$CO, and C$^{18}$O, which are all observable with ALMA.

The raytracing incorporated the observing geometry, set by the distance $d = 100$\,pc, inclination $i=50^\circ$, and position angle $\vartheta=90^\circ$, following the convention where $\vartheta$ is measured E of N to the redshifted side of the major axis. \rev{The inclination can have a pronounced effect on emission surface measurements, as discussed in Section \ref{sec:surfaceuncertainty}.} The {\tt RADMC-3D} raytracing was configured to produce cubes on a spatial grid with 10\,milliarcsecond pixels over a $1024\times1024$ $(\sim10\arcsec)$ field of view and a spectral grid with 50\,m\,s$^{-1}$ channels spanning $\pm5$\,km\,s$^{-1}$ around the systemic velocity ($v_\text{sys} = 0$ here). 

To assess the effects of noise and spatial resolution, we generated two datasets. The first dataset \rev{-- used to test the impact of spatial resolution alone --} was created by convolving the noise-free image cube with a circular Gaussian kernel (FWHM = 0\farcs1). The second dataset \rev{-- which was used to test the combined impact of noise and spatial resolution --} was generated by injecting noise into the pure image cube before applying the same convolution, achieving an rms of 2\,mJy\,beam$^{-1}$ in each 50\,m\,s$^{-1}$ channel. This noise level is typical for current high-quality ALMA datasets, ensuring our synthetic tests better resemble realistic conditions for temperature distribution measurements. \rev{We also discuss the effects of larger beams in Sections \ref{sec:resolution} and \ref{sec:surfaceuncertainty}.}

\section{Methods} \label{sec:methods}
\rev{To constrain the disk's two-dimensional thermal structure, $T(r,z)$, we adopt a simple, data-driven approach that leverages observations with high spatial and spectral resolutions  \citep[as in][]{ Pinte18, Law21, PanequeCarreno23}.} This strategy assumes an axisymmetric structure and kinematics \rev{(i.e., the disk's velocity field)}, and it ignores external contributions to the emission (e.g., clouds or envelope contamination). 

The following section details our approach for extracting and modeling the thermal structure of the disk. To remain agnostic about the radial temperature structure, we conducted all measurements and inferences in independent radial annuli \rev{(each with a width equal to half the beam FWHM)}. In each annulus, we measured the emission elevation and the brightness temperature of each line. From these measurements, we then inferred the vertical temperature structure, \rev{without making any assumptions about the radial temperature behavior}.  \rev{Collating} the results for all the annuli, we constructed a composite constraint on the two-dimensional distribution $T(r, z)$. \rev{We demonstrate this process on both our synthetic image cubes (noise-free and noise-injected, Section \ref{sec:results}) and on archival ALMA observations of HD 163296 (Section \ref{sec:HD163296}.)}

\rev{While this method offers flexibility and computational efficiency, it also requires simplifying assumptions about the disk's emission structure.} The most rigorous way to measure the disk's thermal structure (or any physical property) is to incorporate the process described in Section \ref{sec:model} into a forward-modeling framework. That approach combines a physically motivated parametric structure with radiative transfer calculations to self-consistently infer posterior distributions for all input parameters. Although thorough, this method is computationally demanding: each likelihood evaluation is slow, and the high dimensionality of the parameter space necessitates extensive exploration. \rev{The simplified method is designed to avoid these limitations while still accurately measuring the disk's thermal structure.}

\subsection{Emission Surfaces}
\label{sec:surface}
We began by locating the emission surfaces -- the layers from which emission appears to originate, functioning as the photosphere for each spectral line. These emission heights depend on the excitation conditions in the disk and the integrated optical depths along each line of sight from the observer \rev{\citep{Rosotti25}}.

To measure the elevation of an emission surface, we identified intensity peaks in isovelocity ``loops" observed in channel maps and deprojected them into the disk coordinate frame \citep{Pinte18} using the {\tt disksurf} package \citep{Teague21ds}.
Before conducting the initial extraction, the image cube was smoothed with a 1D Gaussian kernel three times the size of the beam major axis ({\tt smooth=3}), \rev{which improves the signal-to-noise ratio of the data and helps better define the emission peak. We found that this step significantly improved the ability of the code to identify the emission surface.} The initial extraction was clipped at that noise floor ({\tt min\_snr=1}).
Two additional extractions were performed\rev{, using the first extracted surface to define kinematic masks that were convolved with Gaussian kernels having FWHM of 2 and 1 beams  ({\tt nbeams = $[2.0, 1.0]$}). For the second and third extractions, we filtered out pixels with SNR $<$ 2 and SNR $<$ 3 ({\tt min\_snr = $[2, 3]$}), respectively.}  This process yields a set of pixels expected to originate from the disk photosphere (as opposed to the line wings). They can be deprojected to obtain a set of disk-frame coordinates $\{r, z, \delta z\}$, where $r$ is the deprojected radial distance, $z$ is the elevation above the midplane, and $\delta z$ is the uncertainty on the location of each extracted point, assigned to be the ratio of the beam major axis and the pixel signal-to-noise ratio.

Rather than relying on a parametric model to describe this surface, we binned the extracted coordinates into radial annuli $r$ for the next steps.
Since extracting the emission surface coordinates requires knowledge of the disk geometry, these measurements are influenced by the geometric parameters ($i$, $\vartheta$, $x_0$, and $y_0$), where the latter two define the disk center coordinates (set to the origin). \rev{While testing the method, we used the geometric parameters used to generate the model, described in Section \ref{sec:radtran}.}

\subsection{Brightness Temperature Profiles}
\label{sec:Tmethods}
Once the emission surface was identified, we estimated the gas temperatures traced by each spectral line. For an optically thick gas in LTE, the gas temperature is equivalent to the peak brightness temperature along each line-of-sight,
\begin{equation}
\label{eq:Tb}
T_{\rm gas} \equiv T_b = \frac{h\nu}{k_B}\left[\ln\left(\frac{2 h \nu^3}{c^2I_\nu} + 1\right)\right]^{-1}.
\end{equation}
To calculate $T_b$, we constructed a peak intensity (moment-8) map from the cube using the {\tt bettermoments} package \citep{Teague18, Teague19mom} and then used the {\tt gofish} package \citep{Teague19fish} to extract the azimuthally averaged \rev{intensity} along the emission surface. \rev{Then, we converted intensity to brightness temperature using the Planck function as in Equation \ref{eq:Tb}.} To capture uncertainty in the surface elevation, this extraction is repeated over $N$ realizations drawn from the distribution of surface coordinates in each annulus, where $N$ is selected to be of order the number of points extracted with {\tt disksurf}. These measurements can include the entire annulus or be restricted to a specific azimuthal range. This process yields a set of $\{r, T_b, \delta T_b\}$ values representing the radial temperature profile for each spectral line.

\begin{figure*}[ht!]
    \centering
    \includegraphics[width=\textwidth]{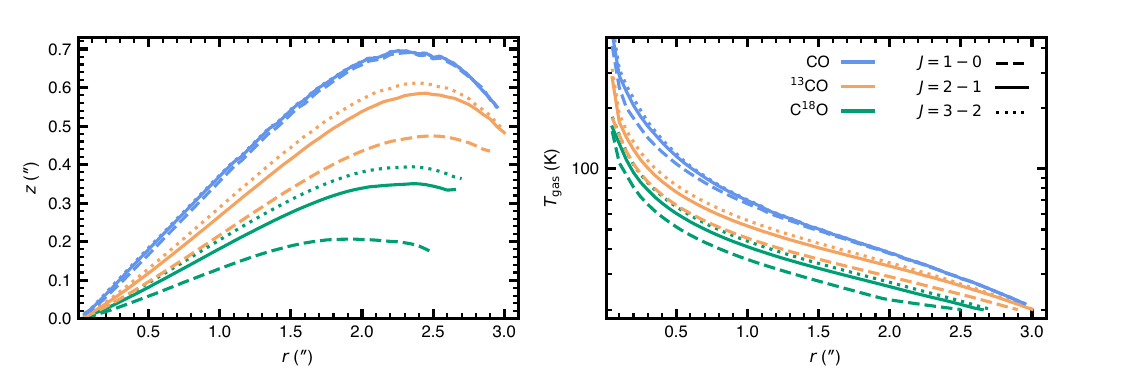}
    \caption{The model $\tau_\text{los} = 2/3$ surfaces ({\it left}) and corresponding radial temperature profiles along those surfaces ({\it right}) for the $J$=1$-$0, 2$-$1, and 3$-$2 emission lines. In this model, the CO emission surfaces are all proximate to the CO photodissociation surface, while the isotopologue surfaces span a range of elevations.}
    \label{fig:all_lines}
\end{figure*}

\rev{Because the moment-8 map selects the peak intensity in each pixel, noise will artificially inflate that value above the true peak intensity along that line-of-sight. To address this, we subtracted the RMS of the cube from the moment-8 maps before extracting the radial profiles, which appears to significantly reduce this bias. A more rigorous approach} would fit a function to the spectrum in each pixel and marginalize over the uncertainties in the parameters, thereby estimating the peak intensity and its uncertainty. \rev{However, radiative transfer effects and the fact that each line of sight intercepts emission from different radii, each at a distinct velocity, prevent a simple spectral model.} In practice, the bias is only significant at large radii, where the signal-to-noise ratio is low. \rev{The magnitude of the bias can be one to several times the RMS of the cube, depending on the line width in the relevant pixel. As a result, the bias becomes significant when $T_b$ is only a few times the noise level.}

\subsection{The $T(z)$ Inference Step}
Once we have measurements of the surface elevation and temperature in each annulus for each line, we could then infer the vertical temperature structure at \rev{an} arbitrary $z$. 
The goal was to recover the injected temperature model defined in Equation \ref{eq:Tannulus} with parameters $\theta = [T_\text{mid}, \gamma, a_z]$ in each annulus.

We defined a Gaussian likelihood function $\mathcal{L} = \p(r, z, T_b, \delta T_b \, | \, \theta)$ where
\begin{equation}
    \ln \mathcal{L} = -\frac{1}{2}\sum_i\left[\frac{(T_{b,i}-T_\text{m}(r, z_{i}))^2}{\delta T_{b,i}^2}\right],
\end{equation}
with the sum running over the set of available optically thick lines in a given annulus.  We used the {\tt emcee} Monte Carlo Markov Chain sampler \citep{ForemanMackey13} to explore the (log-)posterior distribution,
\begin{equation}
    \ln \p(\theta \, | \, \{r, z, T_b, \delta T_b \}) \propto \ln \mathcal{L} + \ln \p(\theta),
\end{equation}
where $\p(\theta)$ denotes the adopted priors.  The posterior sampling was conducted for each of the $N$ realizations of the surface and corresponding $T_b$ and $\delta T_b$, in an effort to propagate the uncertainties on the emission surfaces ($z_i$) into the inferred $T(z_i)$ parameters.

\section{Results}
\label{sec:results}
\subsection{Inferring the Vertical Temperature Structure}
\label{sec:2dinference}

We began with fidelity tests of the vertical \rev{temperature} inference step, setting $z(r)$ and $T_b(r)$ to the true emission location and temperature for each line. We demonstrated that, even with a few emission lines, it is possible to accurately infer the gas temperatures over a range of elevations.

First, we \rev{determined} the true emission surfaces for each line, $z(r)$, \rev{in order to test the inference step without introducing biases that might result from the {\tt disksurf} surface extraction}. Unlike the thermal structure, the location of the emitting region is not explicitly specified; it is instead determined by the radiative transfer process. We used the {\tt tausurf} mode of raytracing in {\tt RADMC-3D} to generate cubes of the disk-frame coordinates ($r, \phi, z$) where the line-of-sight optical depth $\tau_\text{los} = 2/3$. Following the Eddington approximation, this depth corresponds to the photosphere -- where, in optically thick regions, the brightness temperature equals the local gas temperature. Next, for each pixel, we selected the coordinates in the channel where the emission peaks and discarded the rest to avoid the line wings, which are less optically thick and thus originate from lower elevations in the disk. We then adopted radial annuli spaced by 0\farcs05 (half of the Gaussian kernel used to simulate the synthesized beam FWHM) and extending out past the outer edge of the CO-rich layer and averaged the $z$ coordinates within each radial bin. The resulting model emission surfaces for the $J$=1$-$0, 2$-$1, and 3$-$2 lines are shown together in the left panel of Figure \ref{fig:all_lines}. Once the emission surfaces were identified, we evaluated the gas temperatures at those surfaces for each line, resulting in the radial temperature profiles shown in the right panel of Figure \ref{fig:all_lines}. \rev{The $J$=4$-$3 and $J$=6$-$5 emission surfaces follow a similar pattern, and can be found in Appendix \ref{app:additionallines}.}

\begin{figure*}[ht!]
    \centering
    \includegraphics[width=\linewidth]{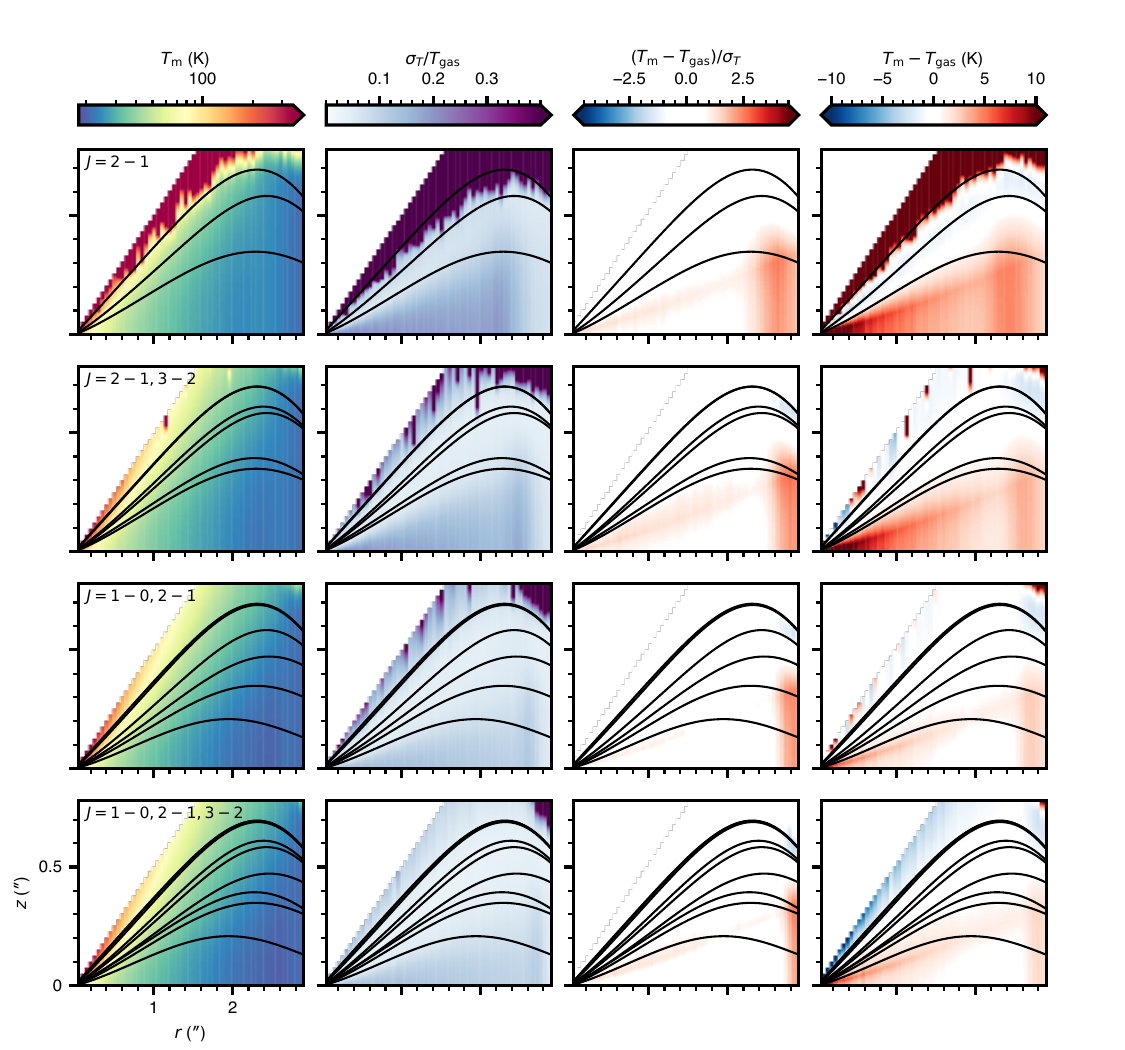}
    \caption{Results of the vertical temperature inference using accurate emission surfaces and radial temperature profiles with nominal uncertainties. The leftmost column shows the median inferred temperatures, the second column shows $\sigma_T/T_\text{gas}$ (precision), the third column shows $\Delta T/\sigma_T$ (accuracy), and the rightmost column shows $\Delta T$. The solid black lines show the emission surfaces included in each inference.}
    \label{fig:verticalinference}
\end{figure*}

After establishing $z(r)$ and $T_b(r)$, we performed the vertical temperature inference in each annulus. To test the precision of the inference, we assumed nominal uncertainties $\delta T = 2$ K and $\delta z = 0.1 z$. These values are larger than the uncertainties obtained by extracting $T$ and $z$ from the noisy image cube (see Sections \ref{sec:surfaceuncertainty} and \ref{sec:realistictest}). We adopted \rev{broad,} uniform priors, 
\begin{equation}
    \text{p}(\theta) = \begin{cases}
        \text{p} (T_\text{mid}) & \sim \mathcal{U}(0, \rev{1500\,K}) \\
        \text{p} (\gamma) & \sim \mathcal{U}(0, 100) \\
        \text{p} (a_z) & \sim \mathcal{U}(0, 1).
    \end{cases}
\end{equation}
\rev{Although more informative priors (for example on $T_\text{mid}$) could enhance accuracy, we deliberately assess the inference’s performance under these extremely broad assumptions.}

We configured an \texttt{emcee} setup with 32 walkers, initialized by random draws from the priors. For each annulus, we sampled the posteriors for 50,000 steps, which was 50 times the autocorrelation time in all cases. We removed the first 10,000 steps as burn-in.

We performed the inference process for various combinations of emission lines. To quantify the performance of the temperature retrievals, we used two metrics: accuracy and precision. We defined accuracy as the deviation of the inferred temperature from the true gas temperature, $|T_\text{m}-T_\text{gas}|$, where $T_\text{m}(r,z)$ is the posterior median at each point in the disk. We defined precision as the uncertainty associated with the retrievals, characterized by the standard deviations of the posterior distributions at each location in the disk, $\sigma_T(r, z)$.

 In Figure \ref{fig:verticalinference}, we present the results for a selection of possible line combinations. Even with $J$=2$-$1 line measurements alone (a common ALMA Band 6 setup), we recovered gas temperatures that are consistent with the true temperatures within the estimated uncertainties.  In the optically thick regions of the disk, the differences between $T_\text{m}$ and the true gas temperature $T_\text{gas}$ are typically less than 1.5 times the posterior standard deviations (i.e., $|T_\text{m} - T_\text{gas}| < 1.5 \sigma_T$). However, $\sigma_T$ can be as high as 20\% of the gas temperatures, even at intermediate elevations between optically thick emission surfaces. \rev{When only one or two measurements are constraining the exponential rise, the temperature constraint in the disk atmosphere is essentially a lower limit, which leads to an extremely high median of the temperature posteriors (as in the case where only $J$=2$-$1 lines are available.)} Introducing $J$=3$-$2 observations improves the precision significantly, shrinking the width of the posteriors within the CO layer to $\sim9\%$. However, the accuracy does not dramatically improve:  $|T_\text{m} - T_\text{gas}|$ is still large in the region below the lowest emission surface.  \rev{The $J$=4$-$3 and $J$=6$-$5 lines behave similarly, as illustrated in Appendix \ref{app:additionallines}.} By contrast, incorporating $J$=1$-$0 observations greatly increases the ability to measure $T_\text{gas}$ throughout the CO layer.

\subsection{Spatial Resolution Bias}
\label{sec:resolution}
\begin{figure}[ht!]
    \centering
    \includegraphics[width=\linewidth]{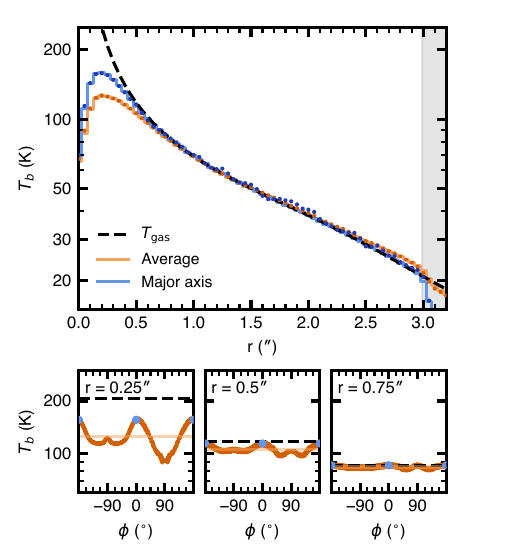}
    \caption{The extracted radial temperature profile (top) and azimuthal temperature profiles in annuli (bottom) for the CO $J$=2$-$1 emission cube. In the top panel, the orange lines and points show the azimuthally averaged brightness temperature, while the blue shows the brightness temperature along the major axis. The lines show extractions from the noiseless cubes, while the points are extracted from the noisy cubes. The black dashed line shows the gas temperature at the $\tau = 2/3$ surface, and the shaded gray region shows where $\tau < 5$. In the bottom panels, the light orange line shows the azimuthal average, and the blue points are along the major axis.}
    \label{fig:resbias}
\end{figure}

Next, we tested the radial temperature profile extraction. Still employing the $\tau=2/3$ surface measured with {\tt RADMC-3D}, we extracted the brightness temperature profile of each line from the cubes \rev{using {\tt gofish}, as described in Section \ref{sec:Tmethods}.} Figure \ref{fig:resbias} shows radial temperature profiles for the CO $J$=2$-$1 line in three scenarios: (1) azimuthally averaged around the entire disk; (2) restricted along only the major axis; and (3) the truth (the injected gas temperature along the $\tau = 2/3$ surface). The lower panels show how the temperature varies with azimuth at three different radii (annuli).

Throughout much of the disk, the extracted brightness temperatures match the true gas temperatures at the $\tau=2/3$ surface. However, at smaller radii ($\lesssim$\,0\farcs8, or 8$\times$ the beam FWHM), the extracted temperature is biased low. In each annulus, the measured $T_b$ along the major axis, where $\phi \approx 0$ or $\pm$180\degr, is closest to the true $T_\text{gas}$.  But $T_b$ decreases toward the minor axes; the ``near" side ($\phi \approx -90$\degr) is brighter than the ``far" side ($\phi \approx +90$\degr).

This intensity bias arises from limited spatial resolution: $T_b \approx T_\text{gas}$ only when the emission fills the beam in the channel where the intensity peaks. If the emitting region at a given frequency is smaller than the beam, the integrated intensity is conserved, but the peak intensity is diluted. \rev{This effect is discussed in detail by \citet{Weaver18}, who suggest using the integrated emission rather than the peak emission map when the beam size is large compared to the channel maps. Although this approach can help remediate the effect of the beam dilution, it can also introduce contamination from the optically thick line wings. We designed an alternate method of correcting the peak intensity measurement.}

Because each channel is convolved with the beam, we analyze a single channel with some velocity $v$. For simplicity, and \rev{because we find that $T_b$ is closest to $T_\text{gas}$ along the major axis (Figure \ref{fig:resbias}),} we will focus on measurements taken within a few degrees of that axis. However, because molecular emission typically arises from an elevated surface, the ``major axis" here is not a simple straight line in the sky-plane. 
In this context, the ``major axis" is in the disk frame and refers to the curve along which the orbital motion is maximally aligned with the line-of-sight. This curve can be computed geometrically by projecting the emission surface $z(r)$ into the sky-plane and identifying the $(x, y)$ locations where the surface is perpendicular to the line-of-sight ($\phi=[-180, 0, 180]^{\circ}$).

\begin{figure*}[ht!]
    \centering
    \includegraphics[width=\linewidth]{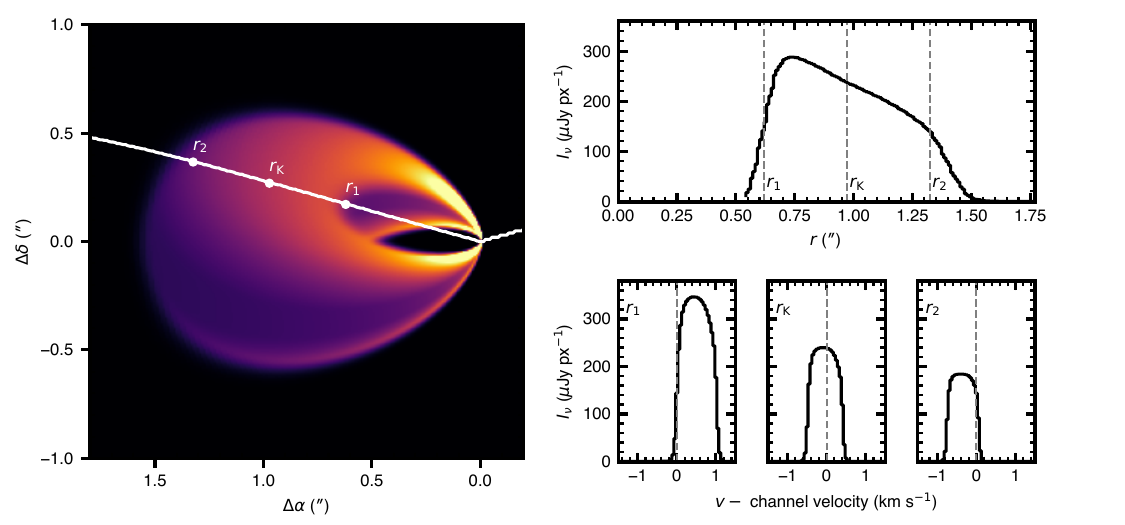}
    \caption{The left panel shows an example full model resolution channel map from the CO $J$=2$-$1 emission cube. The white line shows the major axis. The top right panel shows the radial intensity profile of the front surface along the major axis, with dashed lines at $r_1$ and $r_2$ (where the intensity is half the maximum intensity) and $r_\text{K}$ (the radius where the projected Keplerian velocity is equal to the channel velocity). The three panels on the bottom right show the spectrum of the front surface at the pixels corresponding to $r_1$, $r_\text{K}$, and $r_2$, relative to the velocity of the example channel. The spectrum at $r_\text{K}$ peaks at the channel velocity, while the channel velocity has half the maximum intensity at $r_1$ and $r_2$ (the channel velocity is $v_\text{peak} \pm v_\text{FWHM}/2$).}
    \label{fig:coordinates}
\end{figure*}

To assess whether the emission fills the beam, we consider the full width at half maximum, $\Delta r$, of the emitting region along the major axis in the channel with velocity $v$. The center of the emitting region is the radius $r_\text{K}$, where the projected Keplerian velocity matches the channel velocity $v$.
We define $\Delta r = r_2 - r_1$, where $r_1$ and $r_2$ are the inner and outer boundaries of the region, at which the intensity in that channel falls to half its peak value. Assuming that the temperature gradient across this region is small, the blackbody intensity $B_\nu(T)$ is approximately constant over $[r_1, r_2]$. In that case, the observed intensity in the channel at $r_1$ and $r_2$ is roughly half of the local blackbody intensity, i.e., the optical depth $\tau_{v} = -\ln \frac{1}{2}$.
Figure \ref{fig:coordinates} illustrates this in an example channel map from the CO $J$=2$-$1 cube at full resolution (no beam convolution). The white curve traces the disk major axis. 

The spectrum of each pixel is dictated by the orbital velocity structure and line broadening. In this case, we assume that the line peak at a given radius is Keplerian:
\begin{equation}
\label{eq:vpeak}
v_\text{peak}(r) = v_\text{K}\sqrt\frac{r^3}{(r^2 + z(r)^2)^{3/2}} \sin i.
\end{equation}
where 
\begin{equation}
    v_\text{K}^2 = \frac{GM_*}{r}
\end{equation}
defines the Keplerian velocity.

If we assume that thermal broadening dominates the shape of the line, then the FWHM linewidth is
\begin{equation}
\Delta v(r) = 2 \sqrt{2\ln 2}\, \sqrt{\frac{k_B T_\text{gas}(r)}{m}} 
\end{equation}
where $m$ is the mass of the emitting molecule. Then, since $r_1$ and $r_2$ are the radii where $I_{\nu}$  is half the blackbody intensity,
\begin{equation}
\label{eq:r1}
v_\text{peak}(r_1) - \frac{\Delta v(r_1)}{2} = v
\end{equation}
and
\begin{equation}
\label{eq:r2}
v_\text{peak}(r_2) + \frac{\Delta v(r_2)}{2} = v.
\end{equation}

Rearranging Equations \ref{eq:vpeak}, \ref{eq:r1}, and \ref{eq:r2}, we find the width of the emitting region along the major axis in each channel:
\begin{multline}
    \label{eq:vofr}
    \Delta r(v) = GM_* \sin^2 i \\
    \left[\frac{1}{\left(v - \Delta v(r_2)/2\right)^2}-\frac{1}{\left(v + \Delta v(r_1)/2\right)^2}\right].
\end{multline}

Since we are assuming the temperature gradient is small across $\Delta r$, then $\Delta v(r_1) \approx  \Delta v(r_2) \approx  \Delta v(r_\text{K})$. 
The goal is to correct the extracted $T_b(r)$ profile for this resolution bias. Because $v_\text{peak}$ is the velocity corresponding to the peak intensity at a given $r$, we set $v = v_\text{peak}(r)$ to get the width of the emission as a function of radius and rewrite Equation \ref{eq:vofr}: 
\begin{equation}
    \Delta r(r) = 2GM_* \sin ^2 i \frac{v_\text{peak}(r)\Delta v(r)}{\left(v^2_\text{peak}(r) -\left(\Delta v(r)/2\right)^2 \right)^2}.
\end{equation}

To estimate the effect of beam convolution on the measured brightness, we apply a correction factor to the intensity by assuming a shape for the emitting region. Since the brightness distribution is complicated (top right panel of Figure \ref{fig:coordinates}), we consider two idealized models: a Gaussian profile and a top-hat profile, both centered at $r_\text{K}$. We modeled the beam as a circular Gaussian with FWHM $\sigma_{\rm beam} \times 2 \sqrt{2\ln{2}}$.

If the brightness distribution is approximately a Gaussian with standard deviation $\sigma_r = \Delta r / 2 \sqrt{2 \ln 2}$, the convolved peak intensity is reduced by a factor
\begin{equation}
\label{eq:Iratiogauss}
    \mathcal{K} =\frac{\sigma_r}{\sqrt{\sigma_{\rm beam}^2+\sigma_r^2}}.
\end{equation}
Alternatively, if the brightness distribution were a top hat with width $\Delta r$ such that
\begin{equation}
I_\nu = \begin{cases}
        B_\nu & r_1 < r < r_2 \\
        0 & \text{otherwise},
    \end{cases}
\end{equation}
then the convolved peak intensity is reduced by
\begin{equation}
\label{eq:Iratiotop}
    \mathcal{K} = \text{erf}\left(\frac{\Delta r}{8\sqrt{\ln2}\sigma_\text{beam}}\right).
\end{equation}

In each annulus, for a given model temperature $T_{\rm m}$, we computed the corresponding blackbody intensity $B_\nu(T_{\rm m})$, applied $\mathcal{K}(T_{\rm m})$, and calculated the expected convolved intensity:
\begin{equation}
    I_{\rm m} = \mathcal{K}(r, z, T_{\rm m}, M_*) \ast B_\nu(T_{\rm m}).
\end{equation}
We then compared the model intensity to the measured intensity, $I_\nu$, in that annulus to find the best-fit $T_{\rm m}$. This can be solved with an optimization calculation, but to propagate the uncertainty on the intensity measurement into uncertainties on $T_b$, we forward modeled the temperature. We defined the log-likelihood function $\ln \mathcal{L} = \ln \p(\{r, I, \delta I\} \, | \, T_\text{m})$  as  
\begin{equation}
\label{eq:loglikelihood}
    \ln \mathcal{L} = - \frac{1}{2} \Biggl[ \frac {I -   \mathcal{K}(r,z, T_\text{m}, M_*) \ast B_\nu(T_\text{m})}{\delta I} \Biggr]^2.
\end{equation}

The calculation of $\mathcal{K}$ depends on the elevation of the surface. The same $z(r)$ and $\delta z$ used to extract the temperature profile were also used to perform the correction. The correction also depends on the stellar mass, $M_*$. Hence, to interpret real observations, we need to extract the velocity profile from the data and infer $M_*$ (described in Section \ref{sec:surfaceuncertainty}). For now, we set a $\delta$-function prior on $M_*$ at the input value. 

We sampled the log-posterior distribution of each $T_\text{m}$ draw,
\begin{equation}
    \ln \p(T_\text{m} \, | \, \{r, z_\text{m}, T_b, \delta T_b, M_*\}) \propto \ln \mathcal{L} +\ln \p(T_\text{m})
\end{equation}
as before.
We adopted a broad uniform prior on $T_\text{m}$,\
\begin{equation}
    \p(T_\text{m}) \sim \mathcal{U}(0,\,10,000\,\text{K}).
\end{equation}

\begin{figure}[ht!]
    \centering
    \includegraphics[width=\linewidth]{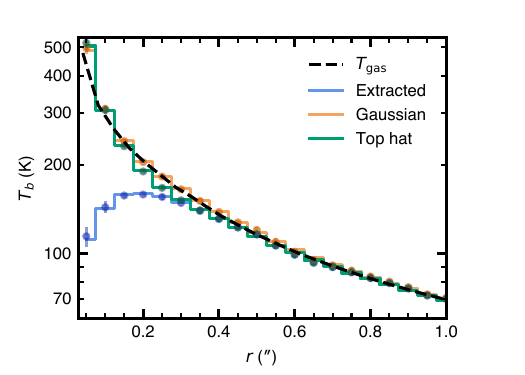}
    \caption{The true gas temperature at the CO $J$=2$-$1 emission surface (black) compared to the brightness temperature extracted from the model image cube (blue) and the extracted brightness temperature corrected according to Equation \ref{eq:Iratiogauss} (green) and Equation \ref{eq:Iratiotop} (orange). The points and uncertainties are extracted from the noisy image cube.}
    \label{fig:Tcorrection}
\end{figure}

For each annulus, we ran 16 {\tt emcee} walkers for 300 steps, initialized by random draws around $T_b$. We found $\langle \tau \rangle \approx 20$, and removed the first 50 ($\sim2\langle\tau\rangle$) steps as burn-in.
In Figure \ref{fig:Tcorrection}, we show the resulting best-fit estimates of the temperature profiles found using the kernels in Equations \ref{eq:Iratiogauss} and \ref{eq:Iratiotop}. Both correction factors gave a significant improvement compared to the extracted $T_b$. For this example, the Gaussian method performed slightly better than the top hat method, but that is not a universal result; it depends on the thermal structure, kinematics, and surface properties of the disk. \rev{Despite the assumptions used in deriving the correction, in practice, the correction is highly accurate even in the presence of very steep temperature gradients ($\frac{dT}{dr} \sim -40$\,K\,au$^{-1}$.)}

This correction is essential not only for accurately measuring gas temperature but also for any extraction of peak intensity profiles from data with limited spatial resolution. It reliably reconstructs the true brightness distribution even with large beams (up to half the disk extent). While the results are most robust beyond the innermost beam, applying this correction consistently yields significantly more accurate profiles than the extracted intensities, even in the central region.

\subsection{Surface and Kinematic Uncertainties}
\label{sec:surfaceuncertainty}

\begin{figure}[ht!]
    \centering
    \includegraphics[width=\linewidth]{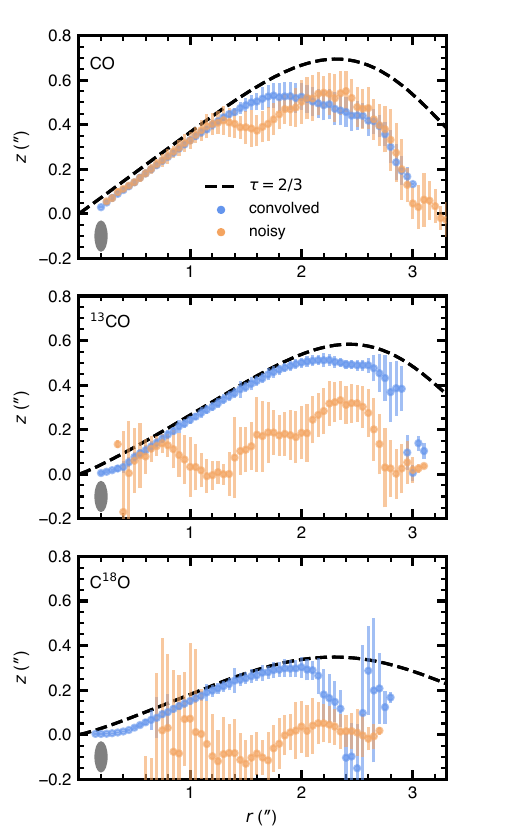}
    \caption{Surfaces extracted using {\tt disksurf} from the $J$=2$-$1 emission cubes of CO, $^{13}$CO, and C$^{18}$O. The black, dashed lines show the true $\tau=2/3$ surface, the blue points show the binned surface points and standard deviations extracted from the convolved, noiseless cube, and the orange points are the same for the convolved, noisy \rev{(rms$~=2$\,mJy\,beam$^{-1}$)} cube. The gray ellipses show the projected beam size \rev{($0\farcs1$ on the sky)}.}
    \label{fig:surfaceposteriors}
\end{figure}

We next examined the downstream effects introduced by uncertainties and biases in the emission surfaces and velocity profiles.  We followed the process described in Section \ref{sec:surface} to extract emission coordinates from the image cubes. Figure \ref{fig:surfaceposteriors} summarizes the extracted surfaces from the $J$=2$-$1 cubes in both the noiseless and noisy cases. Consistent with the findings of \cite{Andrews24}, the extracted coordinates trace the true emission surface in the inner disk, but exhibit a pronounced downward bias at larger radii. \cite{Andrews24} attributed this to two primary reasons: (1) a radiative transfer effect, where the ``near" edge of each half of the isovelocity loop is slightly brighter than the ``far" edge, shifting the apparent emission centroid, and (2) a spatial resolution effect, where it is difficult to distinguish the surfaces when their projected separation is comparable to the beam size. 

In the model we focused on, the first effect is pronounced due to the simulated viewing inclination \rev{($i =50^\circ$)} and the thick CO layer, both of which amplify the asymmetry in the emission structure. \rev{More inclined disks are more significantly impacted, due to a greater difference of optical depth for the ``near" and ``far" sides of the disk.} The resolution effect is especially significant for C$^{18}$O emission, since the front and back surfaces are intrinsically closer together. Adding noise significantly worsens the bias, especially for the $^{13}$CO and C$^{18}$O image cubes, which are inherently fainter than the CO. In the noisy case, for CO $J$=2$-$1, the error in the surface measurement normalized by radius, $\Delta z/r$,  is less than 0.1, and for C$^{18}$O, the measured emission height is indistinguishable from the midplane. 

\rev{The finite beam size can also impact the accuracy of surface extraction: in {\tt disksurf} the {\tt smooth} parameter effectively models this by convolving the brightness distribution with a Gaussian kernel, which both boosts signal‑to‑noise and damps the near–far asymmetry in the isovelocity loops. However, if the resolution approaches the projected separation between the front and back surfaces, the two emission peaks merge and the peak‑finding algorithm can no longer distinguish them, pulling the extracted surfaces artificially together.}

We also tested our ability to extract the kinematic profile, $v(r)$, and to infer the stellar mass, $M_*$, which is required for the temperature correction outlined in Section \ref{sec:resolution}. In each radial bin, we drew 200 samples from the $z$ coordinates in that bin. For each of these surface draws, we extracted the velocity profile from the emission cube following the method of \cite{Andrews24}. We then modeled this velocity profile with Equation \ref{eq:vpeak} to infer $M_*$, excluding the inner $\sim5$ beams to mitigate the spatial resolution bias described by \citet{Andrews24}.
We adopted a uniform prior $M_*$, 
\begin{equation}
    \p(M_*) \sim \mathcal{U}(0, 5\,M_\odot),
\end{equation}
and sampled the posterior with {\tt emcee}, using 32 walkers for 1000 steps. The mean autocorrelation time was $\langle\tau\rangle\approx150$~steps, so we removed the first 300~steps as burn-in.
Although the extracted emission surface showed a significant bias, it had a negligible impact on our estimates of $M_*$ (we found $M_* = 0.99 \pm 0.02M_\odot$). 
\rev{When disk self-gravity and gas pressure are included in calculating the velocity profile of the disk model, the corrected temperature profile underestimates the true temperature by up to 2\%. A more elaborate model that explicitly treats those contributions could correct that error. Such a model might be necessary in the case of more appreciable departures from Keplerian rotation \citep[e.g.,][]{Teague22}.} That model would replace $v_\text{peak}(r)$ in the equations in Section \ref{sec:resolution}.

\begin{figure}[ht!]
    \centering
    \includegraphics[width=\linewidth]{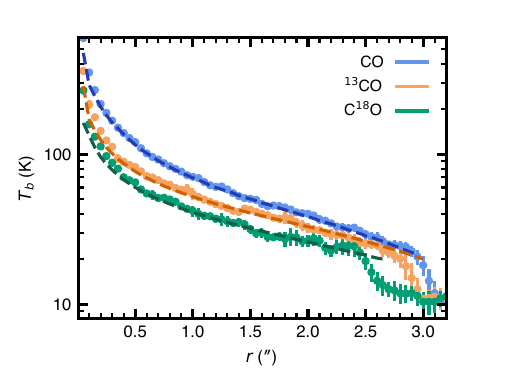}
    \caption{Extracted and corrected temperature measurements for $J$=2$-$1 emission cubes (points) compared to the true gas temperatures at the location of the $\tau=2/3$ surface (dashed lines). The temperature extraction and correction steps were performed using the surfaces extracted from the noisy emission cubes, which are shown in Figure \ref{fig:surfaceposteriors}.}
    \label{fig:finalprofiles}
\end{figure}

Finally, using each of the 200 surface profiles and $M_*$ values, we extracted a radial temperature profile from the peak intensity map. We applied the corrections detailed in Section \ref{sec:resolution} and present the resulting temperature profiles for the $J$=2$-$1 emission lines in Figure \ref{fig:finalprofiles}. Despite the biases in the surfaces, the corrected $T_b(r)$ profiles agree with the expected values to within 10\%\ for $r>10$\,au (outside the central beam).

\subsection{The Complete Inference}
\label{sec:realistictest}

\begin{figure*}[ht!]
    \centering
    \includegraphics[width=\linewidth]{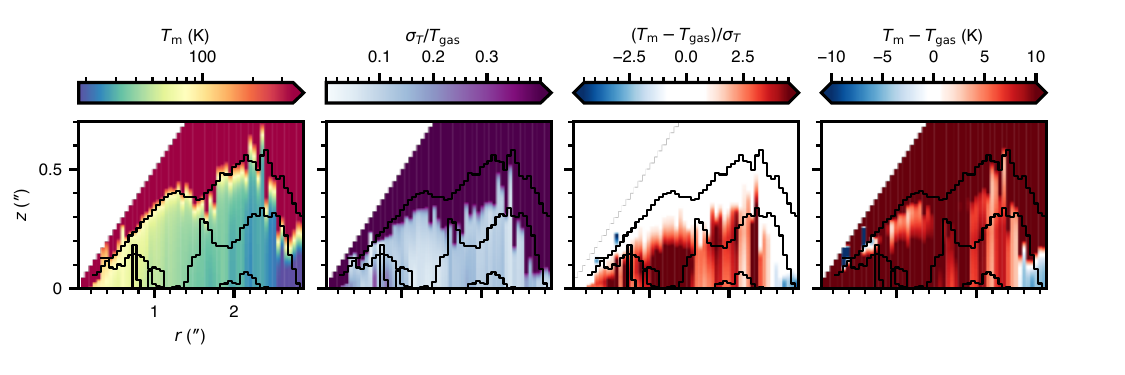}
    \caption{Results of the vertical temperature inference step, using the extracted emission surfaces and temperatures. The black lines show the measured emission surface in each annulus. Panels are as in Figure \ref{fig:verticalinference}.}
    \label{fig:realisticinference}
\end{figure*}

\begin{figure}[ht!]
    \centering
    \includegraphics[width=\linewidth]{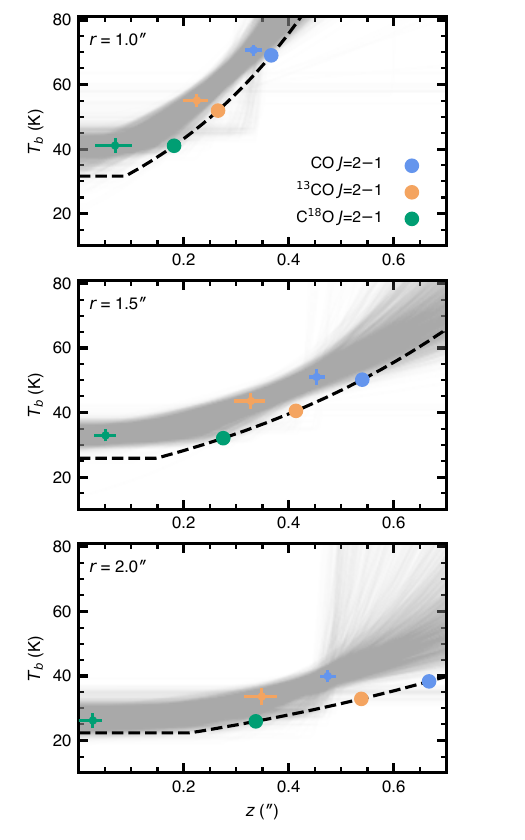}
    \caption{Vertical temperature inference including uncertainties and biases on the temperature and surface elevation measurements. The gray lines show 1000 samples drawn from the posterior fits in each annulus. The black dashed line shows $T_\text{gas}$, and the large colored dots show the true $\tau=2/3$ surface and $T_\text{gas}(r,z)$ for each emission line. The colored points with error bars show the {\it measured} surface elevations and brightness temperatures. The error bars on $z$ and $T$ show the 68\% confidence interval in each bin.} 
    \label{fig:annulifits}
\end{figure}

Using the extracted emission surfaces (Figure \ref{fig:surfaceposteriors}) and radial temperature profiles (Figure \ref{fig:finalprofiles}), we carried out the final inference step to recover the vertical temperature structure; the results are shown in Figure \ref{fig:realisticinference}. Unsurprisingly, the inaccuracies in the extracted surface geometry and (to a lesser extent) the radial temperature profiles propagate into this step, leading to a lower-fidelity reconstruction of the vertical temperature profiles. \rev{Importantly, because the extracted surfaces $z(r)$ are systematically underestimated relative to the true $\tau=2/3$ heights, each temperature measurement (which is itself basically accurate) gets assigned to too-low an elevation. In other words, we end up “overestimating” the temperature at any given elevation simply because we have shifted the entire profile downward.}

Figure \ref{fig:annulifits} \rev{illustrates this issue by showing} the $T(z)$ posteriors for three example annuli. \rev{The posterior median lies above the injected $T_\text{gas}$ at each height, driven by the systematic error in the surface measurement.}
Moreover, as discussed in Section \ref{sec:2dinference}, our method can only reliably constrain temperatures between the highest and lowest emitting layers accessible with the available lines. \rev{As in Figure \ref{fig:all_lines}, above the $^{12}$CO emission surface, the constraint is essentially a lower limit, leading to a very large posterior median. Since the surfaces appear to be lower, the range of accessible elevations appears to be smaller\revii{.}}

\section{Discussion}
\label{sec:discussion}

\subsection{Applications and Limitations}
\label{sec:applications}

Based on these results, we identified several key implications for measuring the two-dimensional thermal  structures of protoplanetary disks.

First, when interpreting peak intensity radial profiles, it is essential to account for a resolution bias that affects the measurements at small radii (Section \ref{sec:resolution}). Ignoring this effect significantly reduces the measured intensities -- and thereby temperatures -- in the disk center. \rev{The bias can extend up to 8 times the size of the beam, and the intensity can be underestimated by as much as 50\%.}

Second, we found that we could infer $T(r,z)$ within the CO layer using only the $J$=2$-$1 transitions of the three most abundant CO isotopologues. Including additional emission lines can improve both precision and accuracy. For the model we studied, adding the $J$=1$-$0 emission from $^{13}$CO and C$^{18}$O significantly improves accuracy and only requires one additional observation. However, in less massive disks with lower CO optical depths, higher energy CO emission lines may offer more useful improvements. For example, the CO $J$=2$-$1 and $J$=6$-$5 lines may then be well separated, in which case $J$=6$-$5 observations would provide a strong lever arm on the vertical temperature structure at high elevations. \rev{\citet{Law23} and \citet{PanequeCarreno23} present surfaces for various CO isotopologues and transitions for a number of disks, showcasing the potential diversity of surface structures and elevation ratios based on the abundance and excitation structure of the disk.} Although the demonstration made here assumes that we know the parametric form of the temperature structure a priori, multiple vertical temperature profiles have been suggested for disks. \rev{For example, the vertical temperature profile that we focus on includes an exponential rise in the atmosphere towards $T=\infty$. In contrast, many works have used temperature profiles that include an isothermal atmosphere above some elevation \citep[e.g.,][]{Dartois03,Dullemond12,Law21}}. Discriminating between them will require additional emission line observations.

Finally, biased emission surfaces are the primary remaining source of error in quantifying the 2-D temperature structures in disks. For an inclined disk, asymmetries in the emission shape confuse the surface extraction significantly, biasing them low by up to $\Delta z/r \approx0.1$ for CO and $\Delta z/r \approx 0.15$ for C$^{18}$O. Although this has a limited effect on the measured radial temperature profiles, it introduces a significant bias when inferring the vertical temperature structure. As a result, improving the fidelity of the surface extraction would substantially enhance our understanding of disk temperatures. However, this remains challenging due to the complexity of radiative transfer effects, for which no simple analytic correction is yet available. \rev{Other surface extraction codes such as {\tt discminer} \citep{Izquierdo21, Izquierdo23} and {\tt alfahor} \citep{PanequeCarreno23} employ slightly different techniques, and as a result may provide some improvement with respect to some of the sources of bias that are evident in the {\tt disksurf} extraction.}

Observationally, measuring 2-D temperature structures in disks requires high sensitivity and resolution. While temperature biases resulting from limited spatial resolution can be corrected, resolving emission surfaces remains a major challenge. Higher spatial resolution is also essential to assess the accuracy of parametric disk models like the one employed in this work, and to determine their reliability in evaluating the inference methods advocated here.

An additional obstacle is determining, annulus by annulus, whether a given transition is optically thick. Two heuristics are commonly used: (i) flagging annuli where brightness temperatures fall below $\sim$20\,K, near the CO freeze-out threshold, and (ii) treating emission beyond the surface turnover as optically thin. The first criterion is independent of the retrieved surface, whereas the second inherits the uncertainties and biases of the surface extraction, and can therefore be overly conservative. Misclassifying annuli as optically thin can result in discarding large amounts of data that could otherwise sharpen the constraints at larger radii.

In this study, we adopted a conservative approach and removed lines beyond the radius where they appeared to become optically thin. Nevertheless, these transitions still encode information: their brightness temperatures set a lower bound on the kinetic temperature at their emission height. Incorporating that limit into the inference, particularly for C$^{18}$O, which turns optically thin at comparatively small radii, could further improve temperature constraints in the outer disk.

\subsection{Fitting Points in Two Dimensions}

\begin{figure*}[ht!]
    \centering   
    \includegraphics[width=\linewidth]{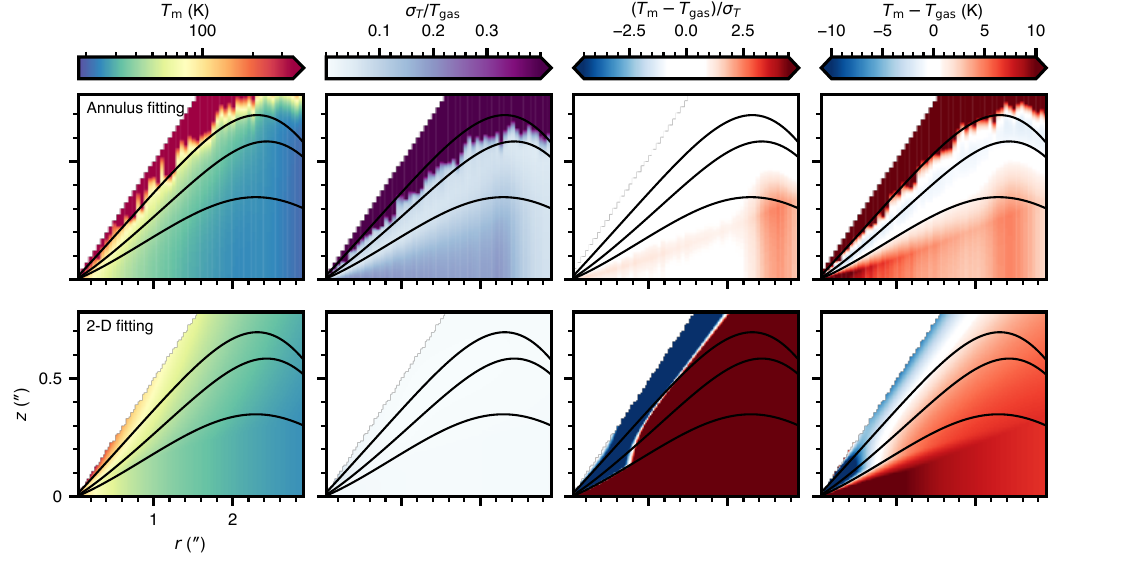}
    \caption{Results of the annulus-by-annulus (top) and 2-D (bottom) temperature inference methods, using the pixels extracted with {\tt disksurf} from the $J$=2$-$1 image cubes and corrected to the $\tau=2/3$ surfaces. The black lines show the emission surface of each line. Panels are as in Figure \ref{fig:verticalinference}.}
    \label{fig:all_points}
\end{figure*}

We described and tested a method for inferring vertical temperature structures in independent radial annuli. An alternative approach, previously adopted by \cite{Law21} and others, involves simultaneously fitting the brightness temperatures over the full set of $\{r, z\}$ points extracted using {\tt disksurf}. This direct 2-D inference method provides a global temperature model, but faces inherent rigidity that introduces biases that can limit its accuracy.

We tested this approach using the synthetic data presented here, following the procedure outlined by \cite{Law21} and summarized here. Each pixel extracted using {\tt disksurf} has an associated $I_\nu$, $\delta I_\nu$, $\nu$, $r$, and $z$. For each of these, we calculated a corresponding $T_b$ and $\delta T_b$. Points with $r<0\farcs8$ were excluded to limit spatial resolution biases near the disk center, and points with $T_b<20$\,K were removed to minimize contamination from optically thin emission. To reduce the effect of the surface measurement bias, we corrected $z$ for each pixel by shifting the set of points in each bin so that the mean value of the bin matched the true surface, while maintaining their scatter.

\rev{While \citet{Law21} used the temperature prescription described in \citet{Dullemond20}, we used the 2-D temperature model described in Equations \ref{eq:Tannulus} and \ref{eq:Tradial} for consistency with the injected temperature structure. This prescription contains} six free parameters: $\{T_{\text{mid},0}, q_\text{mid}, a_{z, 0}, q_a, \gamma_0, q_\gamma\}$. We adopted uniform priors as before,
\begin{equation}
    \p(\theta) = \begin{cases}
    \p (T_\text{mid,0}) & \sim \mathcal{U}(0, 500 K) \\
    \p (q_\text{mid}) & \sim \mathcal{U}(-5, 5)\\
    \p (\gamma) & \sim \mathcal{U}(0, 100) \\
    \p (q_\gamma) & \sim \mathcal{U}(-5,5)\\
    \p (a_z) & \sim \mathcal{U}(0, 1)\\
    \p(q_a) & \sim \mathcal{U}(-5,5).
    \end{cases}
\end{equation}
Posterior distributions were explored with {\tt emcee} using 64 walkers. We initialized the walkers by randomly drawing from the priors. We ran the chains for 10,000 steps, discarding the first 1,000 steps as burn-in.
Figure \ref{fig:all_points} shows the median of the posteriors compared with the injected thermal structure, as well as the results for the annulus-by-annulus inference method. \rev{The 2-D fitting approach yields deceptively tight posteriors, substantially underestimating the true uncertainties at all radii and heights. By comparison, the annulus-by-annulus method more faithfully recovers the midplane temperature, \revii{but sets a lower limit on the atmospheric temperature when only three molecular lines are considered.} However, the 2-D method is subject to multiple sources of bias that the annulus-by-annulus approach avoids.}

First is a sampling bias: the extracted surface points are preferentially drawn from the brighter inner region of the disk, where signal-to-noise is higher and emission surfaces are easier to identify. \rev{However, points in this region are also more likely to be biased in both height and temperature due to limited spatial resolution. Excluding points within some radius helps mitigate this bias, but doing so also reduces the dynamic range of the data, making it more difficult to robustly constrain the temperature structure.}

Second is the intrinsic scatter in the extracted emission surfaces.  As demonstrated in Section \ref{sec:surfaceuncertainty}, there is significant scatter in the elevations of the extracted pixels even though the true $\tau=2/3$ surface is vertically thin. This method assumes that the scatter of $z$ points in each annulus is probing a range of elevations, rather than coming from measurement error. \rev{In reality, the emission for all points extracted at a given radius comes from a single elevation and shares the same brightness temperature.} This results in underestimating the temperature above the emission surface and overestimating the temperature below the surface. \rev{Observational studies \citep[e.g.,][]{Rosenfeld13, Pinte18, Rosotti25} find that the emission structures of real disks are consistent with a vertically thin $\tau=2/3$ layer. While the line wings in each pixel do include contributions from a broader range of elevations, incorporating that information into a two-dimensional temperature structure is challenging due to both radiative transfer effects and geometric projection. The method employed in {\tt disksurf} is designed to extract pixels from the emission peaks in each channel and calculate the elevation of that emission peak. Existing methods are unable to reliably calculate the elevation of pixels drawn from the line wings, but doing so would theoretically enable us to extract the brightness temperature from a wider range of elevations.}

In any case, a more fundamental issue to consider is that we do not have a robust understanding of the radial temperature structures in disks, but the 2-D method requires that a radial functional form be specified. Previous work \rev{\citep[e.g.,][]{Law21, GallowaySprietsma25}} has assumed that the thermal structure follows a power law radially, and hence that each of the parameters in a vertical temperature prescription will follow a radial power law. However, \rev{the observed} radial temperature profiles \rev{presented in those works} appear to diverge from a power law form (e.g., in Section \ref{sec:HD163296}.) Fitting each annulus independently allows flexibility in the radial temperature structure, which enables more empirically oriented measurements of the radial profiles for each parameter.

\subsection{Demonstration with ALMA Data: HD~163296}
\label{sec:HD163296}
After validating the analysis procedure, we applied it to real observations where disk properties are unknown a priori. We selected ALMA observations of the HD~163296 disk from the MAPS project \citep{Oberg21}.  This disk has a \rev{moderate inclination of $i=47^\circ$, which aids in the identification of emission layers, a large radial extent of $\sim 5\farcs0$ in CO} \citep{Law21}, is generally axisymmetric \citep{Teague21}, and does not show problematic foreground contamination. The MAPS data include \rev{sensitive and} high-resolution observations of $^{12}$CO $J$=2$-$1, $^{13}$CO $J$=2$-$1 and $J$=1$-$0, and C$^{18}$O $J$=2$-$1 and $J$=1$-$0.

\begin{figure}[ht!]
    \centering
    \includegraphics[width=\linewidth]{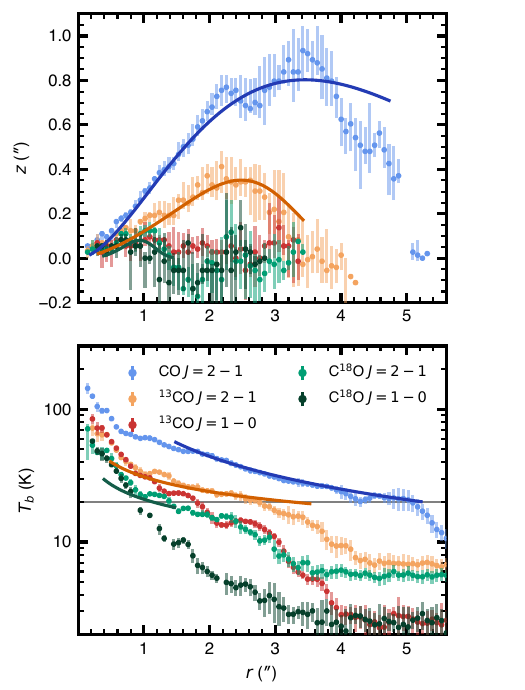}
    \caption{Measured emission surfaces (top) and brightness temperatures (bottom) for HD~163296 binned in annuli. The solid lines show the estimated parametric behaviors for the $J$=2$-$1 lines obtained by \cite{Law21} for reference. \rev{The gray line in the lower panel shows $T_b=20$\,K, below which each emission line was eliminated from the analysis.}}
    \label{fig:HD_163_results}
\end{figure}

We adopted the geometric parameters measured from the same CO $J$=2$-$1 observations in \cite{Izquierdo23}, which are similar to those derived in an independent method by \cite{Teague21}. We extracted the surfaces following the method described in Section \ref{sec:surface} for each emission cube: they are shown together in the top panel of Figure \ref{fig:HD_163_results}, along with the $J$=2$-$1 best fit behaviors derived by \cite{Law21}. For the $J$=1$-$0 image cubes, the emission surface is consistent with $z/r=0$, likely due to the lower resolution of those observations (and the bias on the surface extraction).
We fit the kinematic profile of the $J$=2$-$1 CO observations using the method described in Section \ref{sec:surfaceuncertainty}, and found $M_*=1.93\substack{+0.11\\-0.09}$\,$M_\odot$. We fixed this value for the analyses of all the other cubes. \rev{This value agrees with the range of measurements by \citet{Izquierdo21} and \citet{Teague21}, which vary from 1.9$M_\odot$ to 2.2$M_\odot$. We chose to estimate $M_*$ using the method described above to obtain the closest representation of the velocity profile of our data.}

The brightness temperature profiles extracted and corrected along the \rev{extracted surfaces using the process described in Section \ref{sec:resolution}} are shown in the bottom panel of Figure \ref{fig:HD_163_results}. They deviate from the \cite{Law21} best-fit model profiles, indicating that a single power law is not a good match to the temperature profile. \rev{In addition, the \cite{Law21} profiles are likely significantly biased by the spatial resolution effect in the inner disk.}

\begin{figure}[ht!]
    \centering
    \includegraphics[width=\linewidth]{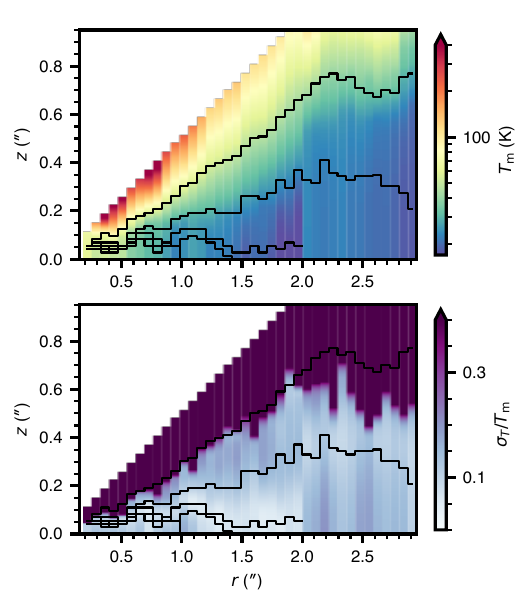}
    \caption{Median values of the temperature posteriors (top) and standard deviation (bottom) for HD~163296. The black lines show the measured emission surface in each annulus.}
    \label{fig:heatmap_hd163}
\end{figure}

\begin{figure}[ht!]
    \centering
    \includegraphics[width=\linewidth]{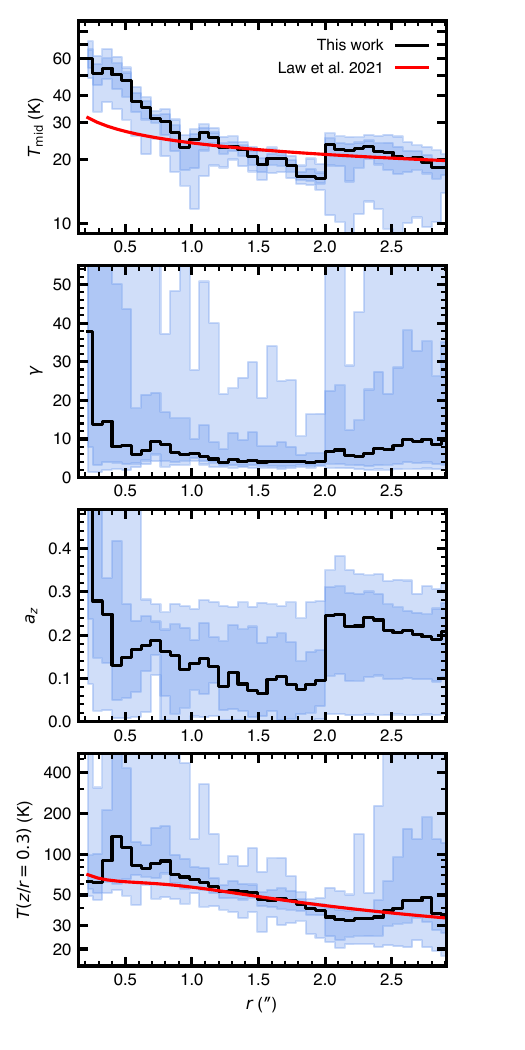}
    \caption{Radial profiles of $T_\text{mid}$, $\gamma$, and $a_z$ for HD~1623926. The black line shows the median of the posteriors, the dark blue shaded region shows the 68\%\ confidence interval, and the light blue shaded region shows the 95\%\ confidence interval. The scatter reflects both measurement uncertainties and  degeneracies in the model fit. The sharp change in the median values exterior to 2\farcs is due to the $^{13}$CO $J$=1$-$0 emission being discarded in these annuli, not because of any discontinuity in the physical structure.}
    \label{fig:parameter_profiles}
\end{figure}
After extracting the radial profiles, we fit the vertical temperature structure in each annulus, using the same functional form and forward modeling set-up described above. We approximated where each line became optically thin based on where $T_b(r) < 20$ K. As a result, we included the C$^{18}$O $J$=1$-$0 measurements out to $r=1\farcs0$, the C$^{18}$O $J$=2$-$1 emission out to $r=1\farcs5$, the $^{13}$CO $J$=1$-$0 emission out to $r=2\farcs0$, and only fit annuli out to $r=3\farcs0$, where at least two lines are available (this is where the $^{13}$CO $J$=2$-$1 emission becomes optically thin).  The resulting thermal structure is shown in Figure \ref{fig:heatmap_hd163}. \rev{The temperature decreases smoothly with radius within $1 \sigma_T$ uncertainties, showing no evidence of substructure. The median temperature profile in each annulus is consistent with the available measurements, but given the small number of points per annulus we cannot rule out alternate temperature prescriptions.}

Figure \ref{fig:parameter_profiles} shows how each of the three fitted parameters \rev{varies with radius, along with the posteriors on the temperature profile in the disk atmosphere, calculated at $z/r =0.3$.} The apparent discontinuity at $r=2\farcs0$ corresponds to the radius beyond which we no longer include $^{13}$CO $J$=1$-$0 in the fit. \rev{Compared to the temperature estimate found in \citet{Law21}, our midplane and atmosphere temperature profiles agree outside $1\farcs0$, but exhibit a steeper temperature increase in the inner disk. This is partially due to the fact that our results appear to diverge from power laws, and partially due to the biases discussed in Section \ref{sec:2dinference}.}

Current constraints are limited by both the surface extraction step and by the relatively small radial extent of optically thick CO isotopologues. Additional measurements of $^{12}$CO, which remains optically thick at larger radii, would help to better constrain the thermal structure beyond 2$\farcs$. For example, the $^{12}$CO $J$=1$-$0 line could be particularly informative, provided the disk's emission morphology follows patterns consistent with the model we tested.

\section{Summary}
\label{sec:summary}
We developed and tested a flexible framework to map the two–dimensional gas temperature structures, $T(r,z)$, in protoplanetary disks from optically thick molecular spectral line emission. We found:

\begin{itemize}
  \item An annulus-by-annulus extraction of emission surface temperatures and elevations, followed by combining a suite of vertical temperature profile inferences, can accurately reconstruct the two-dimensional disk temperatures while effectively propagating uncertainties, even when only CO isotopologue $J$=2$-$1 lines are available.

  \item When spatial resolution is limited, the peak intensity is significantly diluted at small radii. We devise a method to correct the intensity distribution bias by estimating the size of the emitting region, which allows us to produce accurate intensity profiles.

  \item The ultimate precision of the inferred $T(r,z)$ map is set chiefly by the fidelity with which the emission surfaces are recovered; typical biases of $\Delta z/r\approx0.1$–0.15 in the surface location propagate into systematic temperature errors greater than 50\%.

  \item Adding observations of additional transitions expands the vertical coverage and tightens both the accuracy and precision of the temperature constraints across the CO layer. For the high-mass disk model we employed, low energy $J$=1$-$0 lines are particularly useful.

  \item Applying the framework to ALMA observations of HD~163296 recovers a smooth two-dimensional temperature profile with a radial dependence that departs from a simple power law, highlighting the diagnostic power of the flexible annulus-by-annulus method.
\end{itemize}

These results demonstrate that high-quality ALMA data enable reliable, computationally feasible measurements of disk thermal structures.  The framework can be readily applied to other systems and tracers, providing empirical $T(r,z)$ maps that can provide accurate input prescriptions for measuring disk physical properties.

\begin{acknowledgments}
We are grateful to Richard Teague, Stefano Facchini, Viviana Pezzotta, and Andr\'{e}s Izquierdo for thoughtful discussions. A.F. acknowledges support from the National Science Foundation Graduate Research Fellowship under Grant No. DGE 2140743 and from the Smithsonian Institution.

This paper makes use of the following ALMA data: ADS/JAO.ALMA\#2018.1.01055.L. ALMA is a partnership of ESO (representing its member states),
NSF (USA) and NINS (Japan), together with NRC (Canada), MOST and ASIAA (Taiwan), and KASI (Republic of Korea), in cooperation with the Republic of Chile. The Joint ALMA Observatory is operated by ESO, AUI/NRAO and NAOJ. The National Radio Astronomy Observatory is a facility of the National Science Foundation operated under cooperative agreement by Associated Universities, Inc.
\end{acknowledgments}

\software{{\tt astropy} \citep{astropy:2013, astropy:2018, astropy:2022}, {\tt bettermoments} \citep{Teague18, Teague19mom}, {\tt corner} \citep{corner}, {\tt disksurf} \citep{Teague21ds}, {\tt emcee} \citep{ForemanMackey13}, {\tt gofish} \citep{Teague19fish}, {\tt numpy} \citep{numpy}, {\tt matplotlib} \citep{matplotlib}, {\tt RADMC-3D} \citep{Dullemond12}, {\tt SciPy} \citep{scipy}.}

\appendix
\section{CO $J$=4$-$3 and $J$=6$-$5}
\label{app:additionallines}
\begin{figure*}[ht!]
    \centering   
    \includegraphics[width=\linewidth]{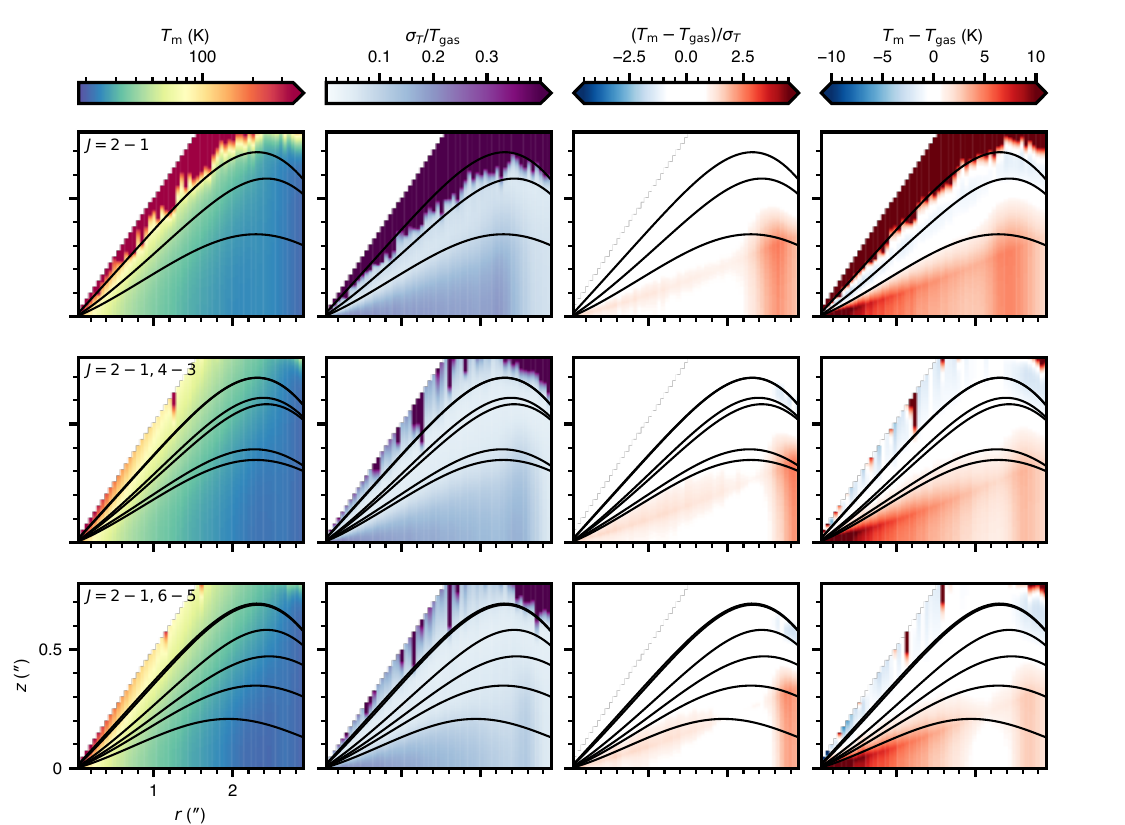}
    \caption{Results of the vertical temperature inference using the $J$=2$-$1, $J$=4$-$3, and $J$=6$-$5 emission lines. Panels are as in Figure \ref{fig:verticalinference}.}
    \label{fig:additionallines}
\end{figure*}
\rev{
In the model structure we focused on, the $J$=2$-$1, $J$=3$-$2, and $J$=4$=$3 emission lines all originate from a similar region in the disk. The $J$=6$-$5 lines of CO and $^{13}$CO also arise from this region, with the C$^{18}$O $J$=6$-$5 emission originating from a slightly lower elevation. This can be attributed to its high excitation temperature combined with the relatively low abundance of C$^{18}$O. As a result, the constraints on the thermal structure obtained by adding the $J$=4$–$3 or $J$=6$-$5 lines to the $J$=2$-$1 dataset follow a similar pattern to the improvements gained by including the $J$=3$–$2 line. That is, there are improvements in accuracy and precision, at high elevations, but the accuracy of the measurement near the midplane is similar even with the additional lines. The $J$=6$-$5 lines provide constraints slightly closer to the disk midplane, but the J$=$1$-$0 lines are still preferred for this goal. These results are shown in Figure \ref{fig:additionallines}.
}
\bibliography{references}{}
\bibliographystyle{aasjournal}

\end{document}